\documentclass[aps,pra,twocolumn,superscriptaddress,amsmath,showpacs,amssymb]{revtex4}
 
\usepackage{bm}
\usepackage{amsmath,amssymb}

\newcommand{\be}{\begin{equation}}
\newcommand{\ee}{\end{equation}}
\newcommand{\ba}{\begin{eqnarray}}
\newcommand{\ea}{\end{eqnarray}}
\newcommand{\bp}{\begin{pmatrix}}
\newcommand{\ep}{\end{pmatrix}}

\newcommand{\nn}{\nonumber}

\newcommand{\Lc}{\mathcal{L}}
\newcommand{\M}{\mathcal{M}}

\bmdefine{\bx}{x}
\bmdefine{\by}{y}
\bmdefine{\bz}{z}
\bmdefine{\bs}{s}
\bmdefine{\bt}{t}

\newcommand{\INT}{\int\!d^3x\:}
\newcommand{\ip}[2]{\Bigl(\:#1,\:#2\:\Bigr)} 

\begin{document}

\title{
Condition for emergence of complex eigenvalues \\
in the Bogoliubov-de Gennes equations}

\author{Y.~Nakamura}
\email{yusuke.n@asagi.waseda.jp}
\affiliation{Department of Materials Science and Engineering,
Waseda University, Tokyo 169-8555, Japan}
\author{M.~Mine}
\email{mine@aoni.waseda.jp}
\affiliation{Department of Physics, Waseda University,
 Tokyo 169-8555, Japan}
\author{M.~Okumura}
\email{okumura.masahiko@jaea.go.jp}
\affiliation{CCSE, Japan Atomic Energy Agency, 6-9-3
Higashi-Ueno, Taito-ku, Tokyo 110-0015, Japan}
\affiliation{CREST (JST), 4-1-8 Honcho, Kawaguchi-shi, Saitama 332-0012,
Japan} 
\author{Y.~Yamanaka}
\email{yamanaka@waseda.jp}
\affiliation{Department of Electronic and Photonic Systems,
Waseda University, Tokyo 169-8555, Japan}

\date{\today}

\begin{abstract}
The condition for the appearance of dynamical instability
of the Bose-condensed system, 
characterized by the emergence of complex eigenvalues in the 
Bogoliubov-de Gennes equations, is studied analytically. 
It is concluded that the degeneracy between a positive-norm eigenmode and a negative-norm one
is essential for the emergence of complex modes. 
Based on the conclusion, we justify the 
two-mode approximation applied in our previous 
work [E. Fukuyama \textit{et al}., Phys.~Rev.~A {\bf 76}, 043608 (2007)], 
in which we analytically 
studied the condition for the existence of complex modes 
when the condensate has a highly quantized vortex. 
\end{abstract}


\maketitle

\section{Introduction}

Experiments of the Bose-Einstein condensates (BECs) 
of neutral atomic gases, first realized in 
1995 \cite{Cornell, Ketterle, Bradley}, have been offering challenging
subjects in theoretical foundations of quantum many-body problem.
Among others, some unstable phenomena of the BECs are very interesting,
since to formulate unstable quantum many-body systems is an open problem.
Observed examples of such phenomena are the split of a doubly quantized
vortex into two singly quantized vortices \cite{Shin}, the decay of a 
condensate flowing in an optical lattice \cite{Fallani}, 
and the decay of the initial configuration in a quenched ferromagnetic
spinor BEC \cite{Sadler}. 
These phenomena are observed at very low temperatures where 
the dissipative mechanism brought by the thermal cloud is negligibly small.
Such instability is called ``dynamical instability," 
and is distinguished from the ``Landau instability," in which the thermal
cloud plays a dissipative role \cite{PS}.

For the theoretical investigations of the instability, 
the Bogoliubov-de Gennes (BdG) equations \cite{Bogoliubov, deGennes, Fetter} 
are employed. 
The BdG equations follow from linearization of the
time-dependent Gross-Pitaevskii (TDGP) equation \cite{Dalfovo} and 
determine the excitation spectrum of the condensate.
The BdG equations have complex eigenvalues for some cases (see below), 
and the presence of the complex eigenvalues is interpreted as 
the sign of dynamical instability.
This instability
is associated with the decay of the initial configuration of the condensate 
and can occur even at zero temperature. 
On the other hand, the Landau instability, 
which is characterized by the negative eigenvalues and in which the thermal cloud drives
the system towards a lower energy state in a dissipative way, is impossible at very
low temperature.
It is reported that the BdG equations have complex eigenvalues in the cases 
where the condensate has a highly quantized vortex \cite{Pu, Kawaguchi}, 
where the condensate flows in an optical lattice \cite{Wu1, Wu2}, 
and where the condensate has gap solitons \cite{Hilligsoe}, 
and in the case of the multi-component BECs \cite{Zhang, Roberts}.
It is suggested that
some degeneracy between a positive mode and a negative one
{\bf in} the BdG equations
is necessary for the emergence of complex eigenvalues \cite{Kawaguchi, Wu1}. 
The analytical investigation on the problem is found 
in Ref.~\cite{Skryabin}, where the instability of vortices in a binary
mixture of BEC is studied in the weak interaction limit.

As for the dynamical instability, another treatment is known, which we refer to as the RK method
given by Rossignoli and Kowalski \cite{RK}. 
In this method the quantum Hamiltonian of the quadratic form of creation and annihilation operators 
is considered, and the complex modes 
appear as a result of diagonalizing it with unusual operators which are neither bosonic nor fermionic ones. 
In the previous work \cite{Fukuyama}, our group has analytically derived 
the condition for the existence of complex modes for the case 
where the condensate has a highly quantized vortex, by using the RK method. 
There a small coupling expansion is adopted, and the two-mode approximation on the Hamiltonian is
assumed and is essential. 
The three-mode analysis is also performed, and it is confirmed that 
the condition for the existence of complex modes is not modified. 
However, it is not clarified why the two-mode approximation is crucial
for the appearance of complex modes. 

In this paper, we study analytically the BdG equations whose two-component eigenfunctions 
are not only of positive-norm but also of negative-norm or of zero-norm.
It is shown that the degeneracy 
between a positive-norm mode and a negative-norm one
is necessary for the emergence of the complex eigenvalues.
To do this, we 
expand the BdG equations in powers of a shift about a value of the coupling constant 
at which all the eigenvalues are real.
The analytic study in Ref.~\cite{Skryabin} has already indicated the importance
of the degeneracy, which was referred to as the frequency resonance of
positive- and negative-energy excitations there.  It is emphasized that 
the present analysis is quite simple and general, not restricted to particular systems. 
Furthermore, our study is based on the expansion of the BdG equations
around a finite (non-zero in general) coupling constant, 
while the one in Ref.~\cite{Skryabin} is limited to the small coupling constant.
We apply the considerations developed here to the case where the condensate has a 
highly quantized vortex, 
and derive the expression of the condition for the emergence of the complex modes 
which completely coincides with our previous one \cite{Fukuyama}. 
Through this analysis, the validity and implication of the two-mode approximation in
our previous work is confirmed.

This paper is organized as follows: 
In Section II, the BdG equations are introduced and their properties are shown.
In Section III, we  expand both the  
GP and BdG equations in powers of a shift of the coupling constant, 
and show that 
the degeneracy between the two modes with different norms causes the complex eigenvalues.
In Section IV, we apply the analysis to the case where the condensate has a highly quantized vortex.
Section V is devoted to the summary.

\section{Bogoliubov-de Gennes Equations and Their Properties}
We start with the following TDGP equation to describe the dynamics of the condensate, 
\be \label{TDGP}
	i\frac{\partial}{\partial t}\Psi(\bx, t) = \left[ K + V(\bx) + g|\Psi(\bx,t)|^2 \right]\Psi(\bx,t) \, ,
\ee
where $K = -\frac{1}{2M}\nabla^2$
with the mass of a neutral atom $M$, and $V(\bx)$ and $g$ are the trap potential for the atoms
and  the coupling constant of self-interaction, respectively.
Throughout this paper $\hbar$ is set to be unity.
The quantity $|\Psi(\bx, t)|^2$ corresponds to the density of condensate atoms, 
and the total condensate number $N$ is given by $N = \INT |\Psi(\bx,t)|^2$.
The stationary state is given as $\Psi(\bx,t) = \xi(\bx) e^{-i\mu t}$ with the chemical potential $\mu$, 
where the function $\xi(\bx)$ satisfies the stationary Gross-Pitaevskii (GP) equation, 
\be 
	\left[ K + V(\bx) - \mu + g|\xi(\bx)|^2  \right] \xi(\bx) = 0 \, .
\ee
To study the excitation spectrum of the condensate, 
let us consider a small fluctuation around the stationary state as
$\Psi(\bx, t) = \left[ \xi(\bx)  + \delta\Psi(\bx,t)\right] e^{-i\mu t}$.
Substituting it into Eq.~(\ref{TDGP}) and linearizing it with respect to 
$\delta\Psi(\bx,t) = u_n(\bx)e^{-i E_n t} + v^*_n(\bx)e^{i E_n^* t}$, 
we obtain the following BdG equations, 
\be
	T \by_n(\bx) = E_n \by_n(\bx) \, . \label{BdG}
\ee  
Here the doublet notation is introduced as
\ba
	\by_n(\bx) &=& \bp u_n(\bx) \\ v_n(\bx) \ep \, , \\ 
	T &=& \bp \Lc & \M \\ -\M^* & -\Lc \ep \, ,
\ea
where 
\ba
	\Lc &=& K + V(\bx) - \mu + 2g|\xi(\bx)|^2 \, ,\\ 
	\M &=& g\xi^2(\bx) \, .
\ea
The eigenvalue $E_n$ is not necessarily real, since $T$ is not a Hermitian operator. 
The similar contents given in this section below are presented in Ref.~\cite{Mine}
for a general case, and in Refs.~\cite{Kawaguchi, Skryabin, Kobayashi} for the
particular systems.

It is easily shown that the operator $T$ has the following algebraic 
properties \cite{Mine, Kobayashi}:
\ba
	\sigma_3 T \sigma_3 &=& T^\dagger \label{sym3} \, ,\\
	\sigma_1 T \sigma_1 &=& -T^*      \label{sym1} \, ,			
\ea
with the $i$-th Pauli matrix $\sigma_i$.
These relations regulate the properties of the eigenfunctions in the following. 

Let us introduce the ``inner product" for arbitrary doublets $\bm{s}(\bx)$ and $\bm{t}(\bx)$ as
\be
	\ip{\bm{s}(\bx)}{\bm{t}(\bx)} = \int\!\!d^3\!x\: \bm{s}^\dagger(\bx) \sigma_3 \bm{t}(\bx) \, ,
\ee
and the squared ``norm" $\| \bm{s} \|^2$ as
\be
	\| \bm{s} \|^2 = \ip{\bm{s}(\bx)}{\bm{s}(\bx)} \, .
\ee
Note that the squared ``norm" may be zero or negative due to the indefinite metric $\sigma_3$.

It is shown from Eq.~(\ref{sym3}) that the operator $T$ is pseudo Hermitian, namely, 
\be	\label{pHermitian}
	\ip{\bm{s}(\bx)}{T \bm{t}(\bx)} = \ip{T \bm{s}(\bx)}{\bm{t}(\bx)} \, .
\ee
Because of the conjugate symmetry of the ``inner product,"
\be \label{conjugate}
	\ip{s}{t}=\ip{t}{s}^* \,, 
\ee
Eq.~(\ref{pHermitian}) 
implies that the matrix element $\ip{\bm{s}}{T\bm{s}}$ is always real.
Furthermore, Eq.~(\ref{pHermitian}) leads to
\be
	(E_n - E_m^{*})\ip{\by_m}{\by_n} = 0 \, ,
\ee
and
\be
	\mathrm{Im}(E_n) \|\by_n\|^2 = 0 \, ,
\ee
where $\by_n$ and $\by_m$ are the eigenfunctions of the BdG equations which
belong to the eigenvalues $E_n$ and $E_m$, respectively.
Two eigenvectors  $\by_n$ and $\by_m$ are orthogonal to each other
for $E_n^* \ne E_m$. 
If $E_n$ is a complex eigenvalue $\left[\mathrm{Im}(E_n)\ne0\right]$, then we have $\|\by_n\|^2=0$.
Equivalently, if $\|\by_n\|^2\ne0$, then $E_n$ is real.

Next, if $\by$ is an eigenfunction belonging to a non-zero real eigenvalue $E$,
one can show from Eq.~(\ref{sym1}) 
that $\bz \equiv \sigma_1 \by^*$ is an eigenfunction belonging to $-E$.
One also finds $\|\by\|^2 = -\|\bz\|^2$, as is seen in 
Refs.~\cite{Wu1, Kawaguchi, Skryabin, Mine, Kobayashi}.
Hereafter, we denote positive- and negative-norm eigenfunctions by $\by_n$ and $\bz_n$,
respectively, which make a pair and are related to each other as $\bz_n=\sigma_1 \by_n^*$. 
Without losing the generality, the absolute values of their norms can be set to unity,
\ba
	\ip{\by_n}{\by_m} &=& \delta_{nm} \, ,\\
	\ip{\bz_n}{\bz_m} &=& -\delta_{nm} \, ,\\
	\ip{\by_n}{\bz_m} &=& 0 \, .
\ea

\section{General Condition for Emergence of Complex Eigenvalues}
In this section, 
in order to derive the condition for the emergence of complex eigenvalues, 
we employ the perturbation theory, dividing the coupling constant $g$ as 
\be
g = g_0 + \varepsilon g' \, , \label{gpert}
\ee
with a sufficiently small parameter $\varepsilon$.
We consider the situation that all the eigenvalues are real for 
the unperturbed coupling constant $g_0$ and that the complex mode 
arises due to the perturbation $\varepsilon g'$ in Eq.~(\ref{gpert}). 
This approach is quite general 
and is applicable to most dynamical instabilities of the BECs
such as the splitting of a vortex, the decay of the flowing
 condensate in an optical lattice.
Similarly to the following analysis, it is also possible to develop 
the perturbation theory not of the 
coupling constant but of the strength of the optical lattice potential
for the flowing condensate in an optical lattice, although we do not present
it in this paper.

The quantities appearing in the GP and BdG equations, (\ref{TDGP}) and
(\ref{BdG}), are expanded in terms of $\varepsilon$ as
\ba
	\xi(\bx) &=& \xi^{(0)}(\bx) + \varepsilon \xi^{(1)}(\bx) + O(\varepsilon^2) \, ,\\
	\by(\bx) &=& \by^{(0)}(\bx) + \varepsilon \by^{(1)}(\bx) + O(\varepsilon^2) \, ,\\
	\mu		 &=& \mu^{(0)}      + \varepsilon \mu^{(1)}      + O(\varepsilon^2) \, ,\\
	E        &=& E  ^{(0)}      + \varepsilon E  ^{(1)}      + O(\varepsilon^2) \, .
\ea
We obtain at the orders of $\varepsilon^0$ and  $\varepsilon^1$, 
\begin{align}	
	\Lc_0 \xi^{(0)} - \M_0 \xi^{(0)*} &= 0 \, , \\
	\Lc_0 \xi^{(1)} - \M_0 \xi^{(1)*} + \Lc' \xi^{(0)} - \M' \xi^{(0)*} &=0\, ,\label{FirstOrderGP}\\
	\left( T_0 - E^{(0)}\right) \by^{(0)}(\bx) &= 0\, ,\\
	\left( T_0 - E^{(0)}\right) \by^{(1)}(\bx) + \left( T'  - E^{(1)}\right) \by^{(0)}(\bx) &=0 \, ,
\end{align}
where
\ba
	\label{0th_order}
	T_0 &=& \bp \Lc_0 & \M_0 \\ -\M_0^* & -\Lc_0 \ep \, ,\\
	\label{1st_order}
	T'  &=& \bp \Lc' & \M' \\ -\M'^* & -\Lc' \ep \, ,
\ea
with
\ba
	\Lc_0 &=& K + V - \mu^{(0)} + 2g_0|\xi^{(0)}|^2 \, ,\\
	\Lc'  &=& 2g'|\xi^{(0)}|^2 + 4g_0 \mathrm{Re}\left( \xi^{(0)} \xi^{(1)*} \right) -\mu^{(1)}\, ,\\
	\M_0 &=& g_0 \xi^{(0)2} \, ,\\
	\M'  &=& g' \xi^{(0)2} + 2g_0 \xi^{(0)} \xi^{(1)} \, .
\ea
Note that the operator $T'$ is pseudo-Hermitian. 

In the case where no degeneracy is present, the first-order eigenvalue is 
\be \label{1st E}
	E^{(1)} = \ip{\by^{(0)}}{T' \by^{(0)}} \, .
\ee
Since $T'$ is pseudo-Hermitian, $E^{(1)}$ is always real 
[see the discussion below Eq.~(\ref{pHermitian})].  
Hence, if all the eigenvalues are real and there is no degeneracy
for the unperturbed coupling constant $g_0$, 
then no complex eigenvalue exists for $g = g_0 + \varepsilon g' $.

Next, let us turn to degenerate cases. 
Here we assume for simplicity that only a single pair of modes is degenerate. 

First, we consider the case in which two positive-norm 
modes $\by_1^{(0)}$ and $\by_2^{(0)}$ share
the same eigenvalue $E^{(0)}$ 
which is real.
Using the degenerate perturbation theory, we obtain the following secular equation which determines $E^{(1)}$,
\be
	\begin{vmatrix} \label{secular_yy}
		\ip{\by_1^{(0)}}{T'\by_1^{(0)}} - E^{(1)}& \ip{\by_1^{(0)}}{T'\by_2^{(0)}} \\
		\ip{\by_2^{(0)}}{T'\by_1^{(0)}} 		 & \ip{\by_2^{(0)}}{T'\by_2^{(0)}} - E^{(1)}
	\end{vmatrix} = 0 \, .
\ee
This implies that $E^{(1)}$ is the eigenvalue of the Hermitian matrix,
\be
	\bp
		\ip{\by_1^{(0)}}{T'\by_1^{(0)}} & \ip{\by_1^{(0)}}{T'\by_2^{(0)}} \\
		\ip{\by_2^{(0)}}{T'\by_1^{(0)}} & \ip{\by_2^{(0)}}{T'\by_2^{(0)}}
	\ep \, ,
\ee
hence $E^{(1)}$ is real.
Likewise, it is also shown that no complex eigenvalue arises from the degeneracy between
two negative-norm modes, $\bz^{(0)}_1$ and $\bz^{(0)}_2$.
  
Suppose that the two modes of one positive- and one negative-norms,
$\by_1^{(0)}$ and $\bz_2^{(0)}$, are degenerate for a real eigenvalue $E^{(0)}$.
We then obtain the secular equation, 
\be
	\begin{vmatrix}
		\ip{\by_1^{(0)}}{T'\by_1^{(0)}} - E^{(1)}& \ip{\by_1^{(0)}}{T'\bz_2^{(0)}} \\
		\ip{\bz_2^{(0)}}{T'\by_1^{(0)}} 		 & \ip{\bz_2^{(0)}}{T'\bz_2^{(0)}} + E^{(1)}
	\end{vmatrix} = 0 \, ,
\ee
which implies that $E^{(1)}$ is the eigenvalue of the non-Hermitian matrix
\be
	\bp
		 \ip{\by_1^{(0)}}{T'\by_1^{(0)}} &  \ip{\by_1^{(0)}}{T'\bz_2^{(0)}} \\
		-\ip{\bz_2^{(0)}}{T'\by_1^{(0)}} & -\ip{\bz_2^{(0)}}{T'\bz_2^{(0)}}
	\ep \, . \label{nonHerMat}
\ee
In this case  $E^{(1)}$ can be complex.
Actually, $E^{(1)}$ becomes complex if the following condition is satisfied, 
\begin{multline} \label{condition}
	\left| \ip{\by_1^{(0)}}{T'\by_1^{(0)}} + \ip{\bz_2^{(0)}}{T'\bz_2^{(0)}} \right| \\
	< 2\left| \ip{\by_1^{(0)}}{T'\bz_2^{(0)}} \right| \, ,
\end{multline}
where Eqs.~(\ref{pHermitian}) and (\ref{conjugate}) are employed.

We conclude that the degeneracy 
between modes of positive- and negative-norms, $\by$ and $\bz$,  
is essential for the emergence of the complex eigenvalue.
This can be extended to the cases where more than two modes are degenerate, 
because the degeneracy between $\by$'s and $\bz$'s gives rise to a non-Hermitian
matrix such as (\ref{nonHerMat}). 
Our conclusion is consistent with the suggestions
in Refs.~\cite{Kawaguchi, Wu1}. 
Note that for the emergence of the complex eigenvalues the degeneracy is necessary 
but not sufficient and that the inequality in Eq.~(\ref{condition}) must be satisfied
in addition.

\section{Application to the Case of a Highly Quantized Vortex and
Validity of Two-Mode Approximation}

In this section, as an application of the formulation developed in the
preceding section, we derive the expression of the condition for the 
emergence of the complex eigenvalues
in a trapped BEC with a highly quantized vortex.
The application elucidates why the two-mode approximation is
valid in our previous work 
where the RK method is used \cite{Fukuyama}.

We consider a harmonic trap potential with a cylindrical symmetry:
\be
	V(r, z) = \frac{1}{2}M(\omega_\perp^2 r^2 + \omega_z^2 z^2) \, ,
\ee 
where $r=\sqrt{x^2 + y^2}$. 
Let us assume that the vortex is created along the $z$-axis.
Then the order parameter $\xi(\bx)$ can be written as
\be
	\xi(\bx) = \sqrt{\frac{N}{2\pi}}e^{i\kappa\theta}f(r,z) \, , 
\ee
with the winding number $\kappa$ which is an integer.
Without losing the generality, $\kappa$ is assumed to be a positive integer.
We consider the case of $g_0=0$, which corresponds to the weak coupling limit. 
The quantities $\Lc_0$, $\M_0$, $\Lc'$, and $\M'$ are written as
\ba
	\Lc_0 &=& K + V -\mu^{(0)} \label{vortex_cL0}, \, \\
	\Lc'  &=& -\mu^{(1)} + 2g'|\xi^{(0)}|^2 \, ,\\
	\M_0 &=& 0 \, ,\\
	\M'  &=& g'\xi^{(0)2} \label{vortex_cM'} \, .
\ea 
The eigenfunctions of the unperturbed BdG equations are written as
\ba
	\by^{(0)}_{n,l,m}(r, \theta, z) 
	&=& \bp U_{n,l,m}(r,\theta,z) \\ 0 \ep \, , \label{by0} \\
	\bz^{(0)}_{n,l,m}(r, \theta, z) 
	&=& \bp 0 \\ U_{n,l,m}^*(r,\theta,z) \ep \, , \label{bz0}
\ea
which belong to the eigenvalues $\epsilon_{n,l,m}$ and 
$-\epsilon_{n,l,m}$, respectively. 
Here, $n\ (=0,1,2,\dots)$ is the principal quantum number, $l\ (=0,\pm1, \pm2, \dots)$ is the magnetic quantum number, and $m\ (=0,1,2,\dots)$ is the quantum number along the $z$ axis. 
The function $U_{n,l,m}(r,\theta,z)$ is expressed as
\be
	U_{n,l,m}(r,\theta,z) = \sqrt{\frac{1}{2\pi}} e^{i(l+\kappa)\theta} u_{n,l,m}(r,z) \, ,
\ee
with the solution $u_{n,l,m}(r,z)$ of the following eigenequation: 
\ba
	&&\biggl[ - \frac{1}{2M} \biggl( \frac{\partial^2}{\partial r^2} +
	\frac{1}{r} \frac{\partial}{\partial r} - \frac{(l+\kappa)^2}{r^2}
	+ \frac{\partial^2}{\partial z^2} \biggr) \notag\\ 
	&&\hspace{2.8cm}+ V(r,z) -\mu^{(0)} \biggr] u_{n,l,m}(r,z) \notag\\ 
	&&=  \epsilon_{n,l,m} u_{n,l,m}(r,z) \, . 
\ea
The explicit form of $u_{n,l,m}(r,z)$ is given by
the Laguerre polynomials $L_n ^k (x)$ and the Hermite polynomials $H_m
(x)$ \cite{LH} as 
\begin{align}
u_{n,l,m} (r,z) & = C_{n,l,m} e^{-\frac{1}{2} \alpha_{\bot}^2 r^2}
(\alpha_{\bot} r )^{|l + \kappa|} L_{n}^{|l + \kappa|}
 (\alpha_{\bot} ^2 r^2) \nn \\  
& \quad {} \times e^{- \frac{1}{2} \alpha_z^2 z^2} H_m (\alpha_z z) \, ,
 \label{un} 
\end{align}
with
\be
C_{n,l,m} = \sqrt{ \frac{2 \, \alpha_{\bot}^2 \alpha_z }{\pi^{1/2}}} 
\cdot \sqrt{ \frac{n!}{2^m \ m! \ (|l + \kappa| + n)!} } \, ,  \label{normc}
\ee
while the eigenvalue is given as
\be
	\epsilon_{n,l,m} = \omega_\perp \left( 2n + |l + \kappa|+ 1\right) 
					 + \omega_z \left( m + \frac{1}{2} \right) - \mu^{(0)}\, .
\ee
The normalization constant (\ref{normc}) is determined so that 
\be
\int\! d^3x \, U^\ast_{n,l,m} (r,\theta,z) \, U_{n',l',m'} (r,\theta,z) 
= \delta_{nn'} \delta_{ll'} \delta_{mm'} \, ,
\ee
and $\| \by^{(0)}_{n,l,m} \|^2 = - \| \bz^{(0)}_{n,l,m} \|^2 = 1$. 

The unperturbed order parameter and the chemical potential are obtained as
\ba
	\xi^{(0)}(r,\theta,z) &=& \sqrt{N} U_{0,0,0}(r,\theta,z) \, ,\\
	\mu^{(0)} &=& \omega_\perp(\kappa + 1) + \frac{1}{2}\omega_z \, .
\ea
Substituting Eqs.~(\ref{vortex_cL0})--(\ref{vortex_cM'}) into Eq.~(\ref{FirstOrderGP}),
we obtain
\be	\label{mu1}
	\mu^{(1)} = D \frac{(2\kappa)!}{(\kappa!)^2} \, , 
\ee
with
\be
	D = \frac{g'N}{\sqrt{8\pi^7}} \frac{\alpha_\perp^2 \alpha_z}{2^{2\kappa + 2}} \, .
\ee

Now, let us evaluate the matrix elements of $T'$,
\begin{widetext}
\ba
	\label{ip_yTy}
	\ip{\by^{(0)}_{n,l,m}}{T' \by^{(0)}_{n',l',m'}} &=& \delta_{ll'} \frac{g'N}{\pi}
	\int\! drdz\: r u_{0,0,0}^2 u_{n,l,m} u_{n',l,m'} - \delta_{nn'} \delta_{ll'} \delta_{mm'} \mu^{(1)} \, ,\\
	\label{ip_zTz}	
	\ip{\bz^{(0)}_{n,l,m}}{T' \bz^{(0)}_{n',l',m'}} &=& \delta_{ll'} \frac{g'N}{\pi}
	\int\! drdz\: r u_{0,0,0}^2 u_{n,l,m} u_{n',l,m'} + \delta_{nn'} \delta_{ll'} \delta_{mm'} \mu^{(1)} \, ,\\
	\label{ip_yTz}
	\ip{\by^{(0)}_{n,l,m}}{T' \bz^{(0)}_{n',l',m'}} &=& \delta_{l,-l'} \frac{g'N}{2\pi}
	\int\! drdz\: r u_{0,0,0}^2 u_{n,l,m} u_{n',-l,m'} \, .
\ea
\end{widetext}
We notice that the matrix elements in Eqs.~(\ref{ip_yTy}) and (\ref{ip_zTz})
vanish unless $l=l'$, while that in Eq.~(\ref{ip_yTz}) vanishes unless $l=-l'$.
According to the general formulation of the degenerate perturbation theory,
the whole secular determinant is a product of secular determinants of
submatrices corresponding to the subspaces spanned by the degenerate eigenfunctions.
We therefore inquire the degeneracy in Eqs.~(\ref{ip_yTy})--(\ref{ip_yTz}). 
Concerning Eqs.~(\ref{ip_yTy}) and (\ref{ip_zTz}),
the degeneracy condition $\epsilon_{n,l,m} = \epsilon_{n',l',m'}$ 
implies $l=l'$, and then $n=n'$ and $m=m'$. 
On the other hand, concerning Eq.~(\ref{ip_yTz}),
the condition is given by $\epsilon_{n,l,m} = -\epsilon_{n',l',m'}$, 
and is satisfied when $l=-l'$, and $n=n'=m=m'=0$ and $|l| \leq \kappa$.
Consequently, only the pairs of $\by_{0,l,0}^{(0)}$ and $\bz_{0,-l,0}^{(0)}$ are degenerate and
give the non-trivial secular determinants.
Thus, the whole secular determinant is a product of the following three kinds of factors:
\begin{widetext}
\begin{eqnarray*}
&\text{(i)}&   \quad  \ip{\by_{n,l,m}^{(0)}}{T' \by_{n,l,m}^{(0)}} - E^{(1)} \, ,\\
&\text{(ii)}&  \quad  \ip{\bz_{n,l,m}^{(0)}}{T' \bz_{n,l,m}^{(0)}} + E^{(1)} \, ,\\
&\text{(iii)}& \quad  \begin{vmatrix}
	\ip{\by_{0,l,0}^{(0)}}{T' \by_{0,l,0}^{(0)}} - E^{(1)} & 
	\ip{\by_{0,l,0}^{(0)}}{T' \bz_{0,-l,0}^{(0)}} \\
	\ip{\bz_{0,-l,0}^{(0)}}{T' \by_{0,l,0}^{(0)}} & 
	\ip{\bz_{0,-l,0}^{(0)}}{T' \bz_{0,-l,0}^{(0)}} + E^{(1)}
	\end{vmatrix} \, .
\end{eqnarray*}
\end{widetext}
It is clear that the only term which can cause $E^{(1)}$ complex is (iii), 
which is the form of the determinant of a $2 \times 2$ matrix, 
as discussed in the preceding section.
This analytic result justifies the two-mode analysis assumed in our previous work \cite{Fukuyama},
and clarifies which two modes are relevant.

Now the condition for the emergence of the complex mode is given by
\begin{multline}
	\left| \ip{\by_{0,l,0}^{(0)}}{T' \by_{0,l,0}^{(0)}} +  
	\ip{\bz_{0,-l,0}^{(0)}}{T' \bz_{0,-l,0}^{(0)}} \right| \\
	<2 \left| \ip{\bz_{0,-l,0}^{(0)}}{T' \by_{0,l,0}^{(0)}} \right| \, .
\end{multline}
Calculating the inner products, 
we finally obtain the expression of the condition as 
\begin{multline}
	\left| \frac{(2\kappa+l)!}{2^l\kappa!(\kappa+l)!} + \frac{(2\kappa-l)!}{2^{-l}\kappa!(\kappa-l)!} 
	- \frac{(2\kappa)!}{(\kappa!)^2} \right| \\
	< \left| \frac{(2\kappa)!}{\kappa!\sqrt{(\kappa+l)!(\kappa-l)!}}\right| \, .
\end{multline}
We can see from the inequality that the complex eigenvalues appear when
 $|l|=2$ with $\kappa=2$, $|l|=2, 3$ with $\kappa=3$, or $|l|=2, 3$ with $\kappa=4$.
In the case of $\kappa=0,1$, the inequality is never satisfied and no complex eigenvalue appears.
This result is consistent with that of the numerical studies 
of the BdG equations \cite{Pu, Kawaguchi}, 
and reproduces our previous one \cite{Fukuyama}. 

In our previous analysis \cite{Fukuyama}, we did not treat the BdG equations but the RK Hamiltonian
with the two-mode approximation, although its justification was not clear then.
In this section, we have confirmed for the system of a highly quantized vortex system that
the two-mode analysis in our previous treatment is justified, as the crucial factor of
the whole secular determinant is the secular determinant of a $2\times 2$ matrix. Furthermore,
it is now clear which two modes are relevant. 

\section{Summary}

We have investigated the condition for the appearance of the dynamical
instability of the Bose-condensed system, 
whose sign is interpreted to be the emergence of the complex eigenvalues in the
BdG equations, using the perturbation theory for both the GP 
and BdG equations with respect to the coupling constant. 
We note in comparison with Ref.~\cite{Skryabin} that our
method is quite simple and general without being confined to 
particular systems, and that the perturbative 
expansion is made around a finite (non-zero in general) coupling 
constant.

Our important conclusion is that the emergence of the complex eigenvalues in the BdG equations
is attributed to the degeneracy between a positive-norm eigenmode and a negative-norm one. 
The expression of the condition for the existence of the complex eigenvalues follows from
 the secular equation of the $2\times 2$ matrix.
We have considered the situation that all the eigenvalues are real for 
the unperturbed coupling constant $g_0$, 
and that the perturbation in the coupling constant gives rise to the complex modes.

Inspired by the analysis on the flowing condensate in an optical lattice
with sufficiently small lattice height \cite{Wu1}, suggesting that 
the complex eigenvalues are caused by the degeneracy, Taylor and Zaremba 
solved the BdG equations in the weak potential limit and
confirmed the suggestion \cite{Taylor}.
Similarly as in the case  of the flowing condensate in an optical lattice, 
the perturbation with respect to the potential strength can be considered.
Following the similar discussion as we developed in this paper, we develop the perturbation
theory not of the coupling constant but of the potential strength. Then it will turn out that
the appropriate degeneracy at the 0-th order of the BdG equations is necessary 
for the emergence of the complex eigenvalues.
In this way, our method presented in this paper includes the method of Taylor and Zaremba.

Furthermore, we have applied the analysis to the case where the condensate 
has a highly quantized vortex, 
and have rederived the expression of the condition for emergence of the complex eigenvalues.
We have confirmed why the two-mode approximation applied in our previous work \cite{Fukuyama} 
is valid, 
deriving the form of the secular equation, which is a product of $2 \times 2$
matrix or less. 

\begin{acknowledgments}
Y.N. is supported partially by 
``Ambient SoC Global COE Program of Waseda University'' 
of the Ministry of Education, Culture, Sports, Science and Technology, Japan.
M.M. is supported partially by the Grant-in-Aid for The 21st
Century COE Program (Physics of Self-organization Systems) at Waseda
University, and Waseda University Grant for Special Research Projects. 
This work is partly supported by Grant-in-Aid for
Scientific Research (C) (No.~17540364) from the Japan
Society for the Promotion of Science and one for Young
Scientists (B) (No.~17740258) from the Ministry of
Education, Culture, Sports, Science and Technology, Japan.
The authors thank the Yukawa Institute for Theoretical Physics at Kyoto University 
for offering us the opportunity to discuss this work during the YITP workshop 
YITP-W-07-07 on ``Thermal Quantum Field Theories and Their Applications''.
\end{acknowledgments}




\begin{thebibliography}{99}
\bibitem{LH}
We use the following conventions:
$L_n^k(x) = e^x \frac{x^{-k}}{n!} \frac{d^n}{dx^n}(e^{-x}x^{n+k})$,
$H_m(x)=(-1)^m e^{x^2} \frac{d^m}{dx^m} e^{-x^2}$.
\end{thebibliography}
\end{document}